%% file: main.tex
\documentclass[conference]{IEEEtran}
\IEEEoverridecommandlockouts
\usepackage{cite}
\usepackage{amsmath,amssymb,amsfonts}
\usepackage{graphicx}
\usepackage{textcomp}
\usepackage{xcolor}
\def\BibTeX{{\rm B\kern-.05em{\sc i\kern-.025em b}\kern-.08em
    T\kern-.1667em\lower.7ex\hbox{E}\kern-.125emX}}

\usepackage{subfigure}
\usepackage[noend]{algpseudocode}
\usepackage{enumitem}
\usepackage{booktabs} 
\usepackage{balance}
\usepackage{algorithm}
\usepackage{subfigure}
\usepackage[noend]{algpseudocode}
\usepackage{enumitem}
\usepackage{hyperref}
\usepackage{comment}
\usepackage{bm}

\begin{document}


\title{Pre-training Recommender Systems via Reinforced Attentive Multi-relational Graph Neural Network}



\author{
    \IEEEauthorblockN{Xiaohan Li\IEEEauthorrefmark{2}, Zhiwei Liu\IEEEauthorrefmark{2}, Stephen Guo\IEEEauthorrefmark{3}, Zheng Liu\IEEEauthorrefmark{2}, Hao Peng\IEEEauthorrefmark{4}, Philip S. Yu\IEEEauthorrefmark{2}, Kannan Achan\IEEEauthorrefmark{3}}
    \IEEEauthorblockA{\IEEEauthorrefmark{2}University of Illinois at Chicago, Chicago, IL, USA
    \\\{xli241, zliu213, zliu212, psyu\}@uic.edu}
    \IEEEauthorblockA{\IEEEauthorrefmark{3}Walmart Global Tech, San Francisco Bay Area, CA, USA
    \\\{SGuo, KAchan\}@walmartlabs.com}
    \IEEEauthorblockA{\IEEEauthorrefmark{4}Beihang University, Beijing, China
    \\penghao@act.buaa.edu.cn}
    
}

\maketitle

\begin{abstract}
Recently, Graph Neural Networks (GNNs) have proven their effectiveness for recommender systems. Existing studies have applied GNNs to capture collaborative relations in the data. However, in real-world scenarios, the relations in a recommendation graph can be of various kinds. For example, two movies may be associated either by the same genre or by the same director/actor. If we use a single graph to elaborate all these relations, the graph can be too complex to process. 
To address this issue, we bring the idea of pre-training to process the complex graph step by step. 
Based on the idea of divide-and-conquer, we separate the large graph into three sub-graphs: user graph, item graph, and user-item interaction graph. Then the user and item embeddings are pre-trained from user and item graphs, respectively.
To conduct pre-training, we construct the multi-relational user graph and item graph, respectively, based on their attributes. 

In this paper, we propose a novel \textbf{R}einforced \textbf{A}ttentive \textbf{M}ulti-relational Graph Neural Network (RAM-GNN) to pre-train user and item embeddings on the user and item graph prior to the recommendation step. Specifically, we design a relation-level attention layer to learn the importance of different relations. Next, a Reinforced Neighbor Sampler (RNS) is applied to search the optimal filtering threshold for sampling top-$k$ similar neighbors in the graph, which avoids the over-smoothing issue. We initialize the recommendation model with the pre-trained user/item embeddings. Finally, an aggregation-based GNN model is utilized to learn from the collaborative relations in the user-item interaction graph and provide recommendations. Our experiments demonstrate that RAM-GNN outperforms other state-of-the-art graph-based recommendation models and multi-relational graph neural networks.
 
\end{abstract}





\input{Sections/introduction}

\input{Sections/model}

\input{Sections/experiments}

\input{Sections/related-works}
\input{Sections/conclusion}

\bibliographystyle{IEEEtran}
\balance
\bibliography{ref} 

\end{document}

%% file: Sections/introduction.tex
\section{Introduction}
With the development of the Internet in recent years, the information explosion problem has become an inevitable issue to consider when designing a large-scale online platform. The overload of information hinders users' ability to find what they really need among a large amount of items. To enhance users' experience, recommender systems have been applied in many web applications including e-commerce \cite{hu2018leveraging,liu2020basconv}, social recommendations \cite{tang2013social,yang2021consisrec}, and movie recommendations \cite{zhang2016collaborative}.


In order to improve performance, Knowledge Graph (KG)-based recommender systems have achieved more consideration recently due to KG's capacity of unifying user-item interactions and their side information in a graph~\cite{wang2019kgat, lu2020meta, wang2018ripplenet, wang2019knowledge, zhao2019intentgc, wang2020reinforced}. However, since the KGs used for recommender systems contain multiple kinds of relations and are generally dense, directly conducting aggregation on KGs would suffer from the over-smoothing issue~\cite{li2018deeper} when learning node embeddings. To be more specific, a GNN model aggregates the neighbor information into the central node. However, too complex neighbor information may contain more noises, which thus impedes the aggregation of a GNN model to retrieve relevant information. 
Also, when we directly apply GNNs on KGs, GNNs are not able to explore high-order connectivity because the number of neighbors grows exponentially w.r.t. the number of layers. Take the movie dataset as an example, when two users share similar ages and two movies have the same director, the connection between user A and movie Y is:
\begin{itemize}
     \item user A $\rightarrow$ age $\rightarrow$ user B $\rightarrow$ movie X $\rightarrow$ director $\rightarrow$ movie Y
\end{itemize}
which is a long path with multi-hop connection that GNNs can hardly explore. 

\begin{figure}
  \includegraphics[width=0.48\textwidth]{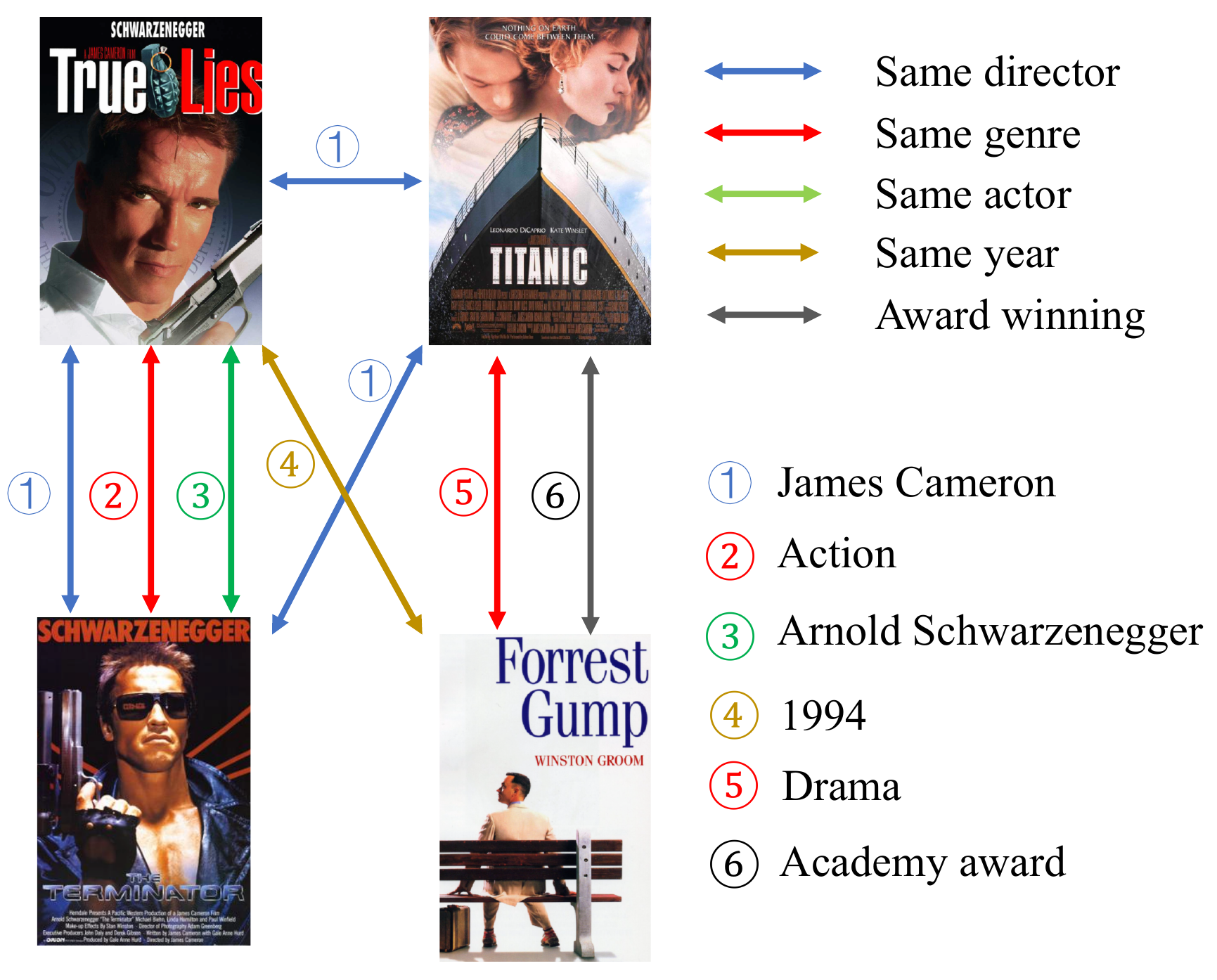}
  \caption{Multi-relational item graph used in the pre-training step. Items sharing the same attributes are connected with an edge. In the graph, item attributes include two parts: relation types (edges) and relation values (numbers). There can be more than one edge between a pair of items.}
  \label{fig:item_graph}
\end{figure}

The spirit of \textit{divide-and-conquer} motivates us to leverage the pre-training and fine-tuning paradigm to learn user/item embeddings. The key idea of pre-training is to learn from other types of data, e.g., data from other sources or data for other tasks, and generate the embeddings~\cite{mikolov2013distributed, devlin2019bert, yang2019xlnet}. In order to preserve the high-order information while avoiding the over-smoothing issue, we first pre-train the model using various user-user and item-item relations in the KG, and then fine-tune it only using the collaborative relation. In the pre-training step, the model can learn from the content of users and items and generate embeddings containing the information of various relations. Moreover, in the fine-tuning step, another model emphasizes on the recommendation task, which only incorporate the collaborative information to achieve the best performance. 

Based on the idea above, we split the long path into several shorter paths which fit the characteristics of GNNs. We divide the complex knowledge graph into three sub-graphs: user-user graph, item-item graph, and user-item interaction graph. In the pre-training step, we construct the multi-relational user graph and item graph, where Figure \ref{fig:item_graph} shows an example of item graph. In these graphs, nodes are connected when they share the same features, so there can be multiple relations between a pair of nodes. In the user-item interaction graph, we utilize the user and item embeddings learned in the pre-training step as the initial embeddings and fine-tune them via a GNN.

However, it is challenging to adopt the aforementioned paradigm on the complex recommendation data. One challenge is \textbf{uneven relation importance}, which means different types of relations in the graph matter unevenly when learning the node embeddings. Figure \ref{fig:item_graph} shows an example of this unevenness. In Figure \ref{fig:item_graph}, there are five types of relations shown in the graph, and some of them are more important when measuring the similarity between movies. For example, the similarity between \textit{True Lies} and \textit{Titanic} is higher than that between \textit{True Lies} and \textit{Forrest Gump}, since a common director is a stronger association than the same year of movies. Therefore, it is required to characterize the importance of those relations. The pre-training model needs to learn the importance score of different relations.


The other challenges is \textbf{uneven distributions}, which indicates the distributions and statistics vary w.r.t. relations. For a particular node, some relations in Figure \ref{fig:item_graph} connect few neighbors for this node, but for some other relations it may have thousands of neighbors. For example, one movie can only have several neighboring movies sharing the same director, but may have millions of movies sharing the same genre. Therefore, the latter relation extremely undermine the uniqueness of the movie when conducting aggregation in the GNN because of the over-smoothing problem. Thus our model should learn from these neighbors appropriately and select the most useful ones to train the model.



To solve the challenges above, we propose 
a novel model named \textbf{R}einforced \textbf{A}ttentive \textbf{M}ulti-relational Graph Neural Network (RAM-GNN). First, we construct a user graph and an item graph with multiple types of relations. Unlike most existing multi-relational graphs \cite{bordes2013translating,schlichtkrull2018modeling, vashishth2019composition}, our user and item graphs have multiple relation between a pair of nodes. Additionally, the relations have values, such as the director of the movie in the "share directors" relation. In this case, we model the meta structure of the multi-relational graph as a quadruplet, which is "item - (relation type) - (relation attribute) - item", to learn the item embeddings as well as the relation embeddings. In RAM-GNN, we apply two modules to solve the two challenges above and learn the node and relation embeddings. One is relation-level attention, which learns the correlations between relations and items. The other is Reinforced Neighbor Sampler (RNS), which samples the top-$k$ similar nodes to filter less relevant neighbors and enhance the quality of learned embeddings via an adaptive Reinforcement Learning (RL) process. To optimize the filtering threshold for each relation, RNS tries to find a trade-off between the average neighbor distances and the total number of the sampled nodes. The model is trained by a GNN loss and a similarity loss jointly with unsupervised learning. We pre-train the RAM-GNN model on the user graph and item graph respectively to learn user and item embeddings. Then another GNN initialized by the pre-trained embeddings is applied to fine-tune the embeddings and provide recommendation results. In experiments, we demonstrate that our model outperforms other state-of-the-art methods. 


In this paper, we summarize our contributions as follows:
\begin{itemize}
    \item We propose the RAM-GNN model to learn the embeddings of users and items for a recommender system. The two main components: relation-level attention and reinforced neighbor sampler provide robust and efficient embedding learning.
    \item We design another GNN to further learn the user and item embeddings based on pre-trained item/user and relation embeddings. This GNN provides a list of items for the user as recommendation results.
    \item We conduct experiments on two real-world datasets for the recommendation task. The experiments demonstrate the effectiveness of the RAM-GNN compared to eight state-of-the-art baselines.
\end{itemize}

%% file: Sections/model.tex
\section{Preliminaries}


\begin{figure*}[tbp]
\includegraphics[width=1\textwidth]{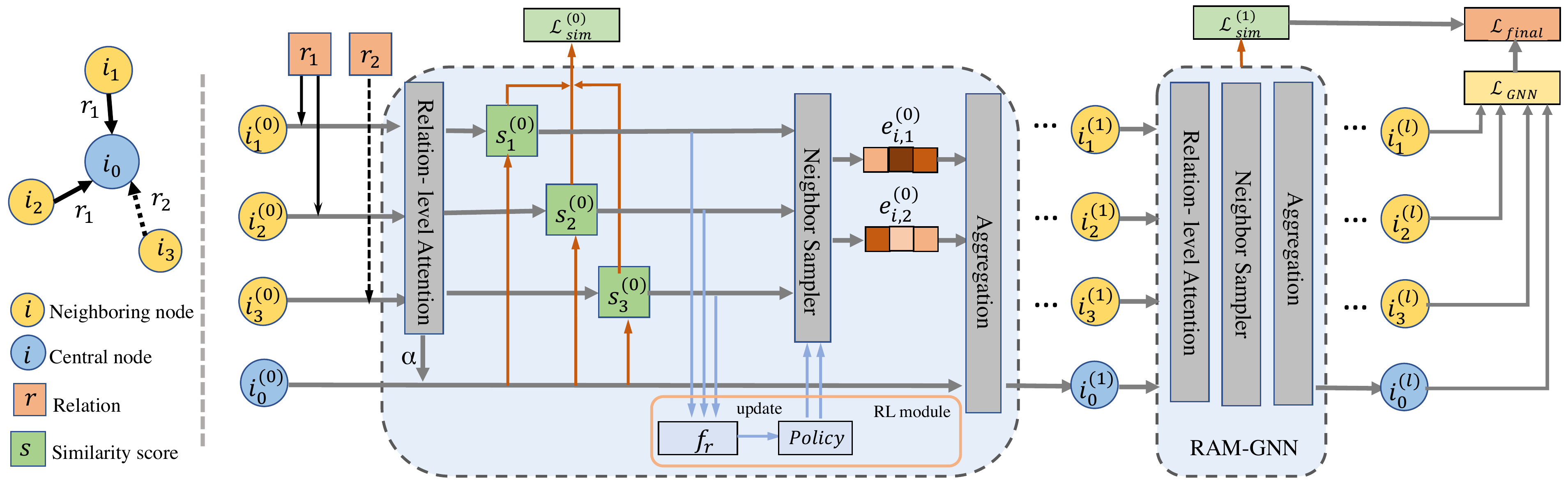}	

\caption{The model structure of RAM-GNN. The central node $i_0$ is connected by neighboring nodes $i_1$, $i_2$ and $i_3$ with relations $r_1$ and $r_2$. Relation-level attention learns the importance of different types of relations. Reinforced Neighbor Sampler finds the most similar neighbors to reduce redundant information and improve efficiency.}
\label{fig:RAM-GNN}
\end{figure*}

Graph Neural Networks~(GNNs) are neural networks that can process graphs directly. By aggregating neighboring information, GNNs can extract structural knowledge from a graph to learn node embeddings. Due to its capacity to process unstructured data, GNNs are widely used on many tasks such as node classification, graph classification, and link prediction. 

The forward propagation procedure of a GNN on graph $G(\mathcal{V},\mathcal{E})$ is to update the representation of each node $v_i \in \mathcal{V}$ through neighboring nodes. Suppose for each node $i$ there is an initial node representation $\bm h^{(0)}_i$ as the input of the model. Then, each hidden layer of GNN learns the central node representation $\bm h^{(l)}_i$ from the previous hidden layer $\bm h^{(l-1)}_i$ by aggregating the neighboring nodes as follows:
\begin{equation}
    \bm h^{(l)}_i = \sigma \Big(\bm W^{(l)} ( \underset{j \in \mathcal{N}(i)}{\bm {Aggr}}(\bm h^{(l-1)}_j) \oplus \bm h^{(l-1)}_i )\Big),
\end{equation}
where $\sigma$ is the non-linear activation function such as LeakyReLU and $\mathcal{N}(i)$ represents the set of neighbors of node $i$ in the graph. 
The neighborhood aggregation function $\bm{Aggr}(\cdot)$  sums neighboring information up and applies an activation function ({\it e.g.}, sigmoid or LeakyReLU). $\oplus$ represents the combination operation of aggregated neighbor embedding and the central node embedding from the previous layer, e.g., concatenation operation.
 


\begin{table}[tb]
  \centering
  \caption{Notations}
  \vspace{2mm}
    \resizebox{0.46\textwidth}{!}{\begin{tabular}{ll}
    \toprule
    Notation                    &Explanation \\
    \midrule
    $\mathbf{e}_i, \mathbf{e}_t, \mathbf{e}_v, \mathbf{e}_j$                 &The embeddings of head item, relation type\\
    &relation value, and tail item\\
    $\mathbf{e}_n$   & The combination embedding of $\mathbf{e}_t$ and $\mathbf{e}_v$ \\
    $\phi$             &Composition operation \\
    $\sigma$       &Activation function\\
    $\mathcal{N}(j)$ & All neighbors of item $j$ \\
    $\bm{Aggr}(\cdot)$ & Aggregation function \\
    $\mathbf{W}_{\text{key}}, \mathbf{W}_{\text{qry}}, \mathbf{W}_{\text{val}}$        &The weight matrices of key,query, and value \\
    $\alpha_{in}$  & The relation-level attention value\\
    $\parallel$    & Concatenation operation\\
    $d(\mathbf{e}_i, \mathbf{e}_j), s(\mathbf{e}_i, \mathbf{e}_j)$    & distance and similarity between node $i$ and $j$ \\
    $\epsilon$ & the change of threshold in each iteration \\
    $AND$   & Average neighbor distance \\
    $\gamma, \Gamma$  &iteration, maximum iteration \\
    $f_r(\cdot)$   & reward function \\
    $\mathcal{L}_{GNN}, \mathcal{L}_{sim}, \mathcal{L}_{final}$ & GNN, similarity and final loss of RAM-GNN \\
    $\mathbf{y}_j,\hat{\mathbf{y}}_j$ & Real value and prediction value \\
    $y_{i,j}$ & The ground truth label of node $i$ and $j$  \\
    $\mathbf{x}_u, \mathbf{x}_i$ & The user and item embedding in recommendation \\
    \bottomrule
    \end{tabular}}
  \label{table:notation}
\end{table}

When the number of edge types is more than one, we call this kind of graph as multi-relational graph. A classic method \cite{schlichtkrull2018modeling} to apply GNN to multi-relational graph is Relational GCN (R-GCN). It re-writes the GNN formula as:
\begin{equation}
    \bm h^{(l)}_i = \sigma \Big( \bm W_r^{(l)} ( \sum_{r \in \mathcal{R}} \underset{ j \in \mathcal{N}_r(i)}{  \bm {Aggr}_{(r)}}( \bm h^{(l-1)}_j) \oplus \bm h^{(l-1)}_i) \Big),
\end{equation}
where $\bm{Aggr}_{(r)},\mathbf{W}_r$ and $\mathcal{N}_r(i)$  denote the aggregation function, weight matrix and the neighbors of node $i$ corresponding to the relation $r$. $\mathcal{R}$ is the set of all relations. However, this method suffers from over-parameterization because each relation is associated with a weight matrix. In this paper, our model tries to solve this problem with entity-relation composition operations to reduce the number of weight matrices in R-GCN to 1.

In the following sections, we first present how to pre-train the item graph with our proposed RAM-GNN. Then, we illustrate how to apply the knowledge learned in the pre-training process to help provide recommendations.

\section{Model}
To overcome the over-smoothing problem and find the most relevant information in the data, we propose the RAM-GNN model to pre-train the user and item embeddings on the user and item graphs, respectively. We use the item graph in Figure \ref{fig:item_graph} as an example to help understand the pre-training paradigm. The user embeddings are pre-trained in an analogous way upon the user graph.

\subsection{Multi-relational Graph} \label{2.2.1}
In real-world applications, there exist multiple relations between a pair of items that have concrete semantics. As shown in Figure~\ref{fig:item_graph}, the relations between items form a multi-relational graph. Given an item pair $(i,j)$, the relations between them are defined as a set of $r=<t,v>$, where $t$ denotes the relation type and $v$ is the relation value. Our goal is to integrate all these relations to learn the item embeddings. Different from the triplets in traditional KGs, the meta structure of the item graph is a quadruplet $(i, t, v, j)$, where $i$ and $j$ are the head and tail item entities, $t$ is the relation type, and $v$ is the relation value. Motivated from methods used in KG embeddings \cite{bordes2013translating}, we propagate the message from tail item $j$ to head item $i$ through relation $r$, and the process is formulated as:
\begin{equation}
    \mathbf{e}_i = \phi (\mathbf{e}_j, \mathbf{e}_t, \mathbf{e}_v),
\label{eq:comp}
\end{equation}
where $\phi$ is a composition operator, $\mathbf{e}_i, \mathbf{e}_t, \mathbf{e}_v$ and $\mathbf{e}_j \in \mathbb{R}^d$ denote the embeddings of head item, relation type, relation value, and tail item respectively in the KG. Different from traditional KG algorithms, which are designed for triplets, our method learns from the quadruplets and composition operations to generate embeddings. For relation types and relations values, we combine them together into integrated embeddings. Here, we concatenate the embeddings of relation types and relations values as:
\begin{equation}
    \mathbf{e}_n = \mathbf{e}_t \parallel \mathbf{e}_v,
\end{equation}
where $\parallel$ is the concatenation operation. In \hyperref[2.3]{Section 2.3}, we will use $\mathbf{e}_v$ in the recommendation step. Then, we can re-write the Eq. \ref{eq:comp} as:
\begin{equation}
    \mathbf{e}_i = \phi (\mathbf{e}_j, \mathbf{e}_n),
\end{equation}
where $\mathbf{e}_n$ is the new embedding that is composed of the relation type $\mathbf{e}_t$ and the relation value $\mathbf{e}_v$. Three kinds of composition operations $\phi$ are applied in our model:
\begin{itemize}
    \item Addition: $\mathbf{e}_i = \mathbf{e}_j + \mathbf{e}_n$,
    \item Multiplication: $\mathbf{e}_i = \mathbf{e}_j * \mathbf{e}_n$,
    \item Circular-correlation: $\mathbf{e}_i = \mathbf{e}_j \star \mathbf{e}_n$.
\end{itemize}
Here, the three operations are motivated from TransE \cite{bordes2013translating}, DistMult\cite{yang2014embedding}, and HolE \cite{nickel2015holographic}. The performance of these operations are discussed in \hyperref[3.5]{Section 3.5.1}.

To learn item embeddings based on the quadruplets, we design our model \textbf{R}einforced \textbf{A}ttentive \textbf{M}ulti-relational Graph Neural Network (RAM-GNN), which are displayed in Figure~\ref{fig:RAM-GNN}. In this model, relation-level and node-level attention layers are proposed to handle the multi-relational graph. Details are introduced in the following sections.

\begin{figure*}
  \centering
  \includegraphics[width=0.85\textwidth]{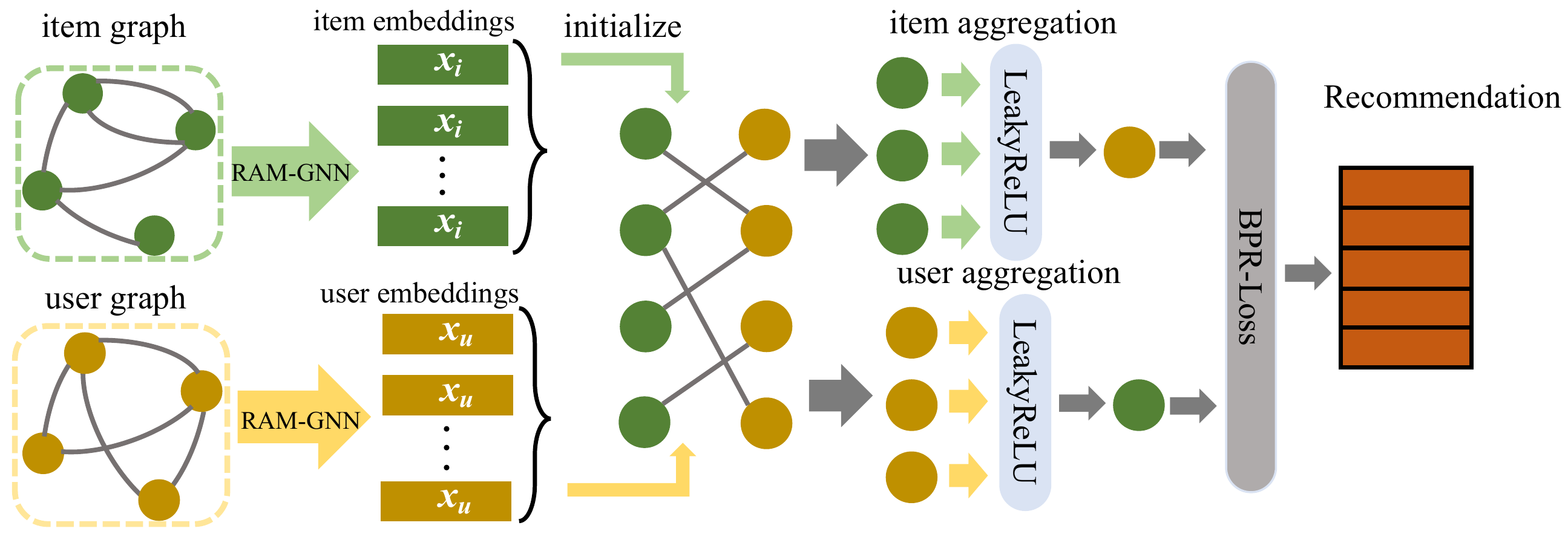}
  \caption{The complete pre-training framework. RAM-GNN pre-trains user and item embeddings on the user and item graph. Then, the learned embeddings are used to initialize the node embeddings in the interaction graph. With another aggregation-based GNN, we obtain the recommendation results through BPR loss.}
 
  \label{fig:framework}
\end{figure*}

\subsection{Relation-level Attention}
For all relations associated with the item $i$, they may contribute unequally in learning the item embedding. For example, the genres and actors of a movie have different importance. Additionally, thousands of movies may share the same genre, while only a few movies may have common actors. To solve this \textbf{uneven relation importance} challenge, we propose the relation-level attention. The attention mechanism \cite{vaswani2017attention} has been widely adopted in existing deep learning models to infer the importance of inputs. In our problem, we adopt the attention mechanism into the multi-relational graph to calculate the attention weights for each relation. We re-write the entity-relation composition operation based on self-attention:
\begin{align}
    \mathbf{a}_{jn} &= \mathbf{p}^T \sigma(\mathbf{W}_{\text{key}} \mathbf{e}_j + \mathbf{W}_{\text{qry}}\mathbf{e}_n + \mathbf{b}), \\
    \mathbf{\alpha}_{jn} &= \text{Softmax}(\mathbf{a}_{jn}), \\
    \mathbf{e}_i &= \mathbf{\alpha}_{jn}\mathbf{W}_{\text{val}}\phi (\mathbf{e}_j, \mathbf{e}_n),
\end{align}
where $\mathbf{p}, \mathbf{b} \in \mathbb{R}^d$, and $\alpha_{jn}$ is the relation-specific attention weight for item pair $(e_i, e_j)$ and relation embedding $e_n$. $\mathbf{W}_{\text{key}}, \mathbf{W}_{\text{qry}}, \mathbf{W}_{\text{val}}$ are key, query, and value weight matrices in the self-attention \cite{vaswani2017attention}, respectively. With relation-specific attention, we can learn how the relation influences the head item and generate the embedding of tail item $j$ considering their correlation.

\subsection{Reinforced Neighbor Sampler}
After the relation-level attention module, we propose the Reinforced Neighbor Sampler (RNS) based on Reinforcement Learning (RL). RNS searches the top-$k$ similar items to reduce redundant information and improve efficiency. In the item or user graph, there can be massive numbers of neighbors for a node. For example, thousands of items can share the same category, and near half of the users are of the same gender. As a result, if we aggregate all neighbors together to learn the embedding of the target node, the unique features of the target node will be overwhelmed in the massive neighbors, which is the over-smoothing issue \cite{li2018deeper}. To deal with \textbf{uneven distributions} challenge, we design an adaptive neighbor sampler to prune the irrelevant neighbors and only select the top-$k$ similar neighboring items to perform aggregation. Given the difference of relations in the graph, the filtering thresholds of neighbor selections are difficult to be pre-defined as hyperparameters. Motivated by \cite{hao2021pre, dou2020enhancing, lai2020policy, peng2021reinforced, wang2020reinforced}, our proposed RNS searches the optimal threshold for each relation via an RL process.

We measure the item similarity based on negative sampling \cite{barkan2016item2vec, wang2020reinforced}. If two items are connected by an edge in the item graph, we take this pair of items as a positive instance. For each item, we also randomly select several irrelevant items as negative samples. We measure the item similarity by minimizing the distance of the positive item pairs and maximizing the distance of negative pairs. Inspired by \cite{he2017neural}, Multi-Layer Perceptron (MLP) is applied to measure the distance of items in RNS. Formally, the distance between two items is as follows:
\begin{equation}
    d(\mathbf{e}_i, \mathbf{e}_j) = \parallel \sigma (MLP(\mathbf{e}_i)) - \sigma(MLP(\mathbf{e}_{j})) \parallel_1
    \label{eq:distance}
\end{equation}
where $\sigma$ is the activation function, and we choose sigmoid in the above equation. $i$ is the target item, and $j$ is the positive or negative samples w.r.t. $i$. The output of the $MLP$ layer is a scalar. With the sigmoid activation, the distance is projected to $(0,1)$, where a closer distance represents two items are stronger similar. To measure the similarity of a pair of items and simplify calculation, we convert distances to similarity scores as:
\begin{equation}
    s(\mathbf{e}_i, \mathbf{e}_j) = 1 - d(\mathbf{e}_i, \mathbf{e}_j),
\end{equation}
where $s(\mathbf{e}_i, \mathbf{e}_j)$ is the similarity score of item $i$ and $j$, and a higher similarity score means $i$ and $j$ are more similar. We use the similarity score to discriminate whether the two items are similar or not.

Based on the item similarity, we define a cross-entropy based similarity loss as the loss function of MLP layers. The similarity loss is a supplement of the GNN loss, which is formulated as:
\begin{equation}
    \mathcal{L}_{sim} = \sum_{j \in \mathcal{I}} -\log y_{i,j} \cdot s(\mathbf{e}_i, \mathbf{e}_j),
\end{equation}
where $\mathcal{I}$ is an item set containing all positive and negative samples w.r.t. item $i$, and $y_{i,j}$ is the ground truth to determine whether $j$ is a positive or negative sample. With the similarity loss, we can learn the similarity score for each item pair and rank them decreasingly to find the most similar ones.

In the item graph, there are multiple types of relations. The number of items we select to conduct aggregation should vary w.r.t. different relations. Therefore, we adopt filtering thresholds for all relations. However, incorporating those thresholds as hyper-parameters is time-consuming because the number of relations can be extremely large.

To improve efficiency in the training process, we propose Reinforced Neighbor Sampler (RNS) to automatically search for the optimal filtering threshold for each type of relation during the training process. We express the RL process of finding filtering thresholds as Bernoulli Multi-armed Bandit (BMAB) $\mathcal{B}(\mathcal{A},\mathcal{R},\mathcal{T})$ \cite{vermorel2005multi}, 
which is a simplified version of Markov decision process as there is no state in it. 
In the BMAB, $\mathcal{A},\mathcal{R}$ and $\mathcal{T}$ represent the action space, the reward function, and the terminal condition, respectively. Initially, we set the $k_t$ as the filtering thresholds of the neighbor sampler corresponding to a relation type $t$. Then we utilize the reward function $f_r$ to determine the action of increasing or decreasing the $k_t$. Specifically, the BMAB process are defined as follows:

\textbf{Action space.} The action represents how we adjust the $k_t$ based on the reward function. We define a fixed small value $\epsilon$ as the action. $k_t$ is increased or decreased by $\epsilon$ to find the optimal filtering thresholds.

\textbf{Reward function.} We design the reward function to determine increasing or decreasing the $k_t$ based on the distance of items we define in Eq. \ref{eq:distance}. The reward function discovers the most similar 
items iteratively to achieve the minimum Average Neighbor Distance (AND) with as large a filtering threshold as possible. The AND in iteration $\gamma$ is calculated as follows:
\begin{equation}
    AND^{\gamma} = \frac{\sum_{j\in \mathcal{N}_{k_t}(i)}d(\mathbf{e}_i, \mathbf{e}_j)^\gamma}{|\mathcal{N}_{k_t}(i)|},
\end{equation}
where $\mathcal{N}_{k_t}(i)$ is the top $k_t$ neighbor items of item $i$. It is a trade-off between the AND and filtering threshold $k_t$, since we intend to have more neighbors into aggregation and less AND. Based on this idea, we can adjust the AND in the reward function as:
\begin{equation}
    f_r(AND^{\gamma}) = \left\{
\begin{aligned}
+1 & , & AND^{\gamma-1} \geq AND^{\gamma}, \\
-1 & , & AND^{\gamma-1} < AND^{\gamma},
\end{aligned}
\right.
\end{equation}
where $f_r$ is the reward function. If the output of $f_r$ is positive, $k_t$ is increased by $\epsilon$ and vice versa. In this way, a positive output of $f_r$ leads to a smaller AND compared to last iteration so that we can increase $k_r$ to have more neighbors involved. In opposite, if the $f_r$ is negative, we need to decrease $k_r$ to limit the AND. These two actions take place back and forth until reaching a convergence. 

\textbf{Termination condition.} The RL process converges when the AND does not vibrate explicitly during the training iteration, which leads to the termination condition as:
\begin{equation}
    \big| \sum_{\gamma-10}^\gamma f_r(AND^\gamma) \big| \leq \epsilon, \text{       where } \gamma >10.
\end{equation}
As this inequality represents the cumulative reward of the recent ten training iterations is less than $\epsilon$, we claim the RL converges under this termination condition. When the termination condition has reached, we fix the filtering threshold in the following GNN training iterations.

\subsection{RAM-GNN}
To combine the relation-level attention and reinforced neighbor sampler into a unified framework, we propose our Reinforced Attentive Multi-relational Graph Neural Network (RAM-GNN) to generate node embeddings. After the RNS module, we select the top-$k$ similar neighbors for each item, so the next step is to aggregate all the filtered neighbors to learn the item embeddings. Since we have taken the relation type into account in the relation-level attention module, the aggregation process is defined as:
\begin{equation}
    \mathbf{e}^{(l)}_i = \sigma \Big( \mathbf{\alpha}_{jn}\mathbf{W}^{(l)}_{\text{val}} \big( \underset{j\in \mathcal{N}_{k_t}(i)}{\bm{Aggr}} \big(\phi (\mathbf{e}^{(l-1)}_j, \mathbf{e}^{(l-1)}_n) \big) \oplus \mathbf{e}^{(l-1)}_i \big) \Big),
\end{equation}
where $l$ is the number of layer in GNN and $j$ is the selected neighbors from RNS. The aggregation layer can be stacked layer-to-layer and construct a deep neural network.

The loss function of the RAM-GNN is composed of two parts: GNN loss and similarity loss. The GNN loss serves for training all parameters in the model, while the similarity loss is mainly designed for helping RNS. The GNN is also trained via negative sampling, and the loss function is:
\begin{equation}
    \mathcal{L}_{GNN} = \sum_{i \in \mathcal{V}} \sum_{j \in \mathcal{I}} -\log y_{i,j} \cdot s(\mathbf{e}_i,\mathbf{e}_j),
\end{equation}
where $\mathcal{V}$ is the set of all nodes in the graph as we need to go through the whole graph to generate node embeddings. The final loss function of RAM-GNN is the combination of GNN loss, similarity loss, and regularization terms, whose formula is:
\begin{equation}
    \mathcal{L}_{final} = \mathcal{L}_{GNN} + \lambda_\gamma \sum_{\gamma = 1}^\Gamma \mathcal{L}_{sim}^\gamma + \lambda_1 \|\Theta_1\|,
\end{equation}
where $\lambda_\gamma, \lambda_1$ are weight hyperparameters and $\Theta$ is all parameters in the RAM-GNN. $\Gamma$ is the iteration number when the RL process converges. All weight matrices in the the pre-training model are randomly initialized and learned in an end-to-end back-propagation training paradigm.

\subsection{Recommendation}\label{2.3}
After learning the pre-trained item embeddings, we further leverage the user-item interaction graph to fine-tune the user/item embeddings via a GNN. Motivated by \cite{wang2019neural}, we learn the user/item embeddings through layer-to-layer aggregation to incorporate the collaborative relation. Different from previous models, we utilize the learned relation embeddings of the items from the pre-training process. In the aggregation process, we concatenate the relation value embeddings together with the item embeddings to learn a better representation of items. 

The structure of the recommendation framework is shown in Figure \ref{fig:framework}. The first-order propagation aggregates the item embeddings into user embeddings:
\begin{align}
    \mathbf{x}^{(0)}_i&= \mathbf{W}_3(\mathbf{e}_i \parallel \mathbf{e}_v),   \\
    \mathbf{x}^{(l)}_u &= \sigma \Big( \mathbf{W}^{(l)}_4 ( \underset{i \in \mathcal{N}(u)}{\bm{Aggr}}(\mathbf{x}^{(l-1)}_i) \oplus \mathbf{x}^{(l-1)}_u) \Big), \label{eq:user}
\end{align}
where $\mathbf{x}_i, \mathbf{x}_u \in \mathbb{R}^d$ represent item and user embeddings, respectively. $\mathbf{W_3} \in \mathbb{R}^{d \times d'}, \mathbf{W_4} \in \mathbb{R}^{d \times d}$ are both trainable weight matrices and $d'$ is the length of concatenated embeddings. $\parallel$ is the concatenation operation to combine the relation value embeddings with the pre-trained item embeddings. Note that the number of $\mathbf{e}_v$ may be more than one. As such, we concatenate the embeddings of all relation values. With first-order propagation, the user embeddings can incorporate the embeddings of interacted items and their attributes.

The second-order propagation aggregates the user embeddings into item embeddings. The process is similar to the item side aggregation, which is:
\begin{align}
    \mathbf{x}^{(0)}_u&= \mathbf{W}_3(\mathbf{e}_u \parallel \mathbf{e}_v),   \\
    \mathbf{x}^{(l)}_i &= \sigma \Big( \mathbf{W}^{(l)}_4 ( \underset{i \in \mathcal{N}(u)}{\bm{Aggr}}(\mathbf{x}^{(l-1)}_u) \oplus \mathbf{x}^{(l-1)}_i) \Big), \label{eq:item}
\end{align}
In this way, we can further fine-tune the pre-trained embeddings with the collaborative relation to make the embeddings more comprehensive. After the second-order propagation, Eq. \ref{eq:user} and Eq. \ref{eq:item} can be stacked interchangeably to build a deeper neural network and explore higher-order proximity.

Based on user and item embeddings, we utilize a widely used BPR loss \cite{rendle2012bpr} to train the model. The objective function is as follows:
\begin{align}
    \hat{y}_{u,i} &= \mathbf{x}_i^T \mathbf{x}_u \\
    \mathbb{L}(y_{u,i},\hat{y}_{u,i}) &= - \ln{\sigma (y_{u,i} - \hat{y}_{u,i})} + \lambda_2 \|\Theta_2\|,
\end{align}
where $\Theta_2$ is all trainable parameters in the GNN model, $\|\Theta_2\|$ is the $L_2$ normalization of all trainable parameters, and $\lambda$ is the regularization coefficient. The recommendation process is also trained by the back-propagation algorithm.

%% file: Sections/experiments.tex
\section{Experiments}
In this section, we conduct experiments on two real-world datasets to evaluate the performance of our proposed RAM-GNN model and the whole pre-training framework. We aim to answer the following research questions:
\begin{itemize}
    \item \textbf{RQ1}: How does the whole pre-training framework perform compared with other recommender system algorithms?
    \item \textbf{RQ2}: How does RAM-GNN perform compared with state-of-the-art multi-relational GNN models?
    \item \textbf{RQ3}: What is the influence of the relation-level attention and reinforced neighbor sampler?
    \item \textbf{RQ4}: How does RNS perform in finding the optimal filtering threshold?
    \item \textbf{RQ5}: How do we select suitable hyperparameters (e.g., choices of composition operations, embedding dimensions) when training the model?
\end{itemize}

\subsection{Experimental Settings}
\subsubsection{Datasets}
We conduct our experiments on two public datasets: MovieLens\footnote{https://grouplens.org/datasets/movielens/} and KKBox\footnote{https://www.kaggle.com/c/kkbox-music-recommendation-challenge/data}.

\begin{table}[]
    \centering
    \caption{The number of users, items, interactions, and relation types in the datasets. `user rel.' and `item rel.' represent the number of relation types of users and items, respectively.}
    \resizebox{0.47\textwidth}{!}{\begin{tabular}{cccccc}
    \toprule
         Datasets & Items & Users & Interactions & User Rel. & Item Rel.  \\
    \midrule
        MovieLens & 1,682 & 943 & 100,000 & 3 & 4 \\
        KKBox & 24,613 & 61,877 & 2,170,690 & 2 & 4 \\
    \bottomrule
    
    \end{tabular}}
    
    \label{tab:dataset}
\end{table}

\begin{itemize}
    \item \textbf{MovieLens.} It is a widely used benchmark dataset published by GroupLens \cite{harper2015movielens}. The user relation types we used in this dataset are ages, genders and occupations. For item relation types, we use genres, directors, movies, and released years. The graph we build to pre-train the item graph is similar to Figure \ref{fig:item_graph}. The user ratings are transformed into binary numbers to indicate implicit feedback. We utilize users' ages and occupations to construct the user graph.
    
    \item \textbf{KKBox. } This dataset was first introduced in the WSDM Cup 2018 Challenge\footnote{https://wsdm-cup-2018.kkbox.events}. It contains four types of relations, including genres, artists, composers and lyricists. We also use ages and living cities as user features.
\end{itemize}

The statistics of these two datasets are listed in Table \ref{tab:dataset}.

\begin{table*}[t]
  \centering
  \caption{Experiments on two datasets comparing our proposed RAM-GNN framework with eight baseline models using the metrics: Hit Rate (HR), Mean Reciprocal Rank (MRR), and Normalized Discounted Cumulative Gain (NDCG). The bold and underlined numbers indicate the best and second-best results on each dataset and metric, respectively. "Improvement" means the minimum improvement among all baselines.}
    \resizebox{1\textwidth}{!}{\begin{tabular}{clccccccccccc}
    \toprule
    {Datasets}  
  &\multicolumn{6}{c}{MovieLens}  &\multicolumn{6}{c}{KKBox}  \\
  \cmidrule(lr){1-1} \cmidrule(lr){2-4} \cmidrule(lr){5-7}  \cmidrule(lr){8-10} \cmidrule(lr){11-13} 
   {Models} & HR@10 & MRR@10 & NDCG@10 & HR@20 & MRR@20 & NDCG@20 & HR@10 & MRR@10 & NDCG@10 & HR@20 & MRR@20 & NDCG@20 \\
    \midrule 
      MF-BPR   &  0.125 & 0.042  & 0.061 & 0.205 & 0.046 & 0.079 & 0.663 & 0.404 & 0.465 & 0.763 & 0.411 & 0.489\\
      FM & 0.143 & 0.051  & 0.072 & 0.208 & 0.055 & 0.087 & 0.701 & 0.426 & 0.493 & 0.799 & 0.432 & 0.515   \\
      NFM   &  0.146 & 0.054  & 0.077 & 0.213 & 0.057 & 0.089 & 0.717 & 0.441 & 0.512 & 0.795 & 0.443 & 0.523 \\
      FISM  &  0.132  & 0.049  &0.068  & 0.209 & 0.053   & 0.087 & 0.696 & 0.410 & 0.484 & 0.764 & 0.426 & 0.524   \\
      CKE & 0.140 & 0.048 & 0.069 & 0.209  & 0.053 & 0.088 & 0.693 & 0.433 & 0.495 & 0.787 & 0.439 & 0.531   \\
      LightGCN   &   0.151 & 0.056  & 0.079 & 0.224 & 0.058 & 0.094 & 0.745 & 0.539 & 0.568 & 0.821 & 0.510 & 0.546\\
      KGPolicy  & 0.154 & 0.058 & \underline{0.083} & 0.231 & 0.062 & \underline{0.104} & 0.778 & 0.562 & 0.591 & 0.847 & 0.554 & 0.613 \\
      RCF   &  \underline{0.159} & \underline{0.059}  & 0.082 & \underline{0.235}   & \underline{0.064} & 0.102 &   \underline{0.794} & \underline{0.572}  & \underline{0.625} & \underline{0.856}   & \underline{0.576} & \underline{0.641}\\
      RAM-GNN & \textbf{0.164} & \textbf{0.062}  & \textbf{0.087} & \textbf{0.242} & \textbf{0.066} & \textbf{0.107} &  \textbf{0.812} & \textbf{0.584}  & \textbf{0.647} & \textbf{0.873} & \textbf{0.589} & \textbf{0.660}\\
     \midrule
     Improvement & 4.46\% & 5.08\% & 4.82\% & 2.98\% & 3.13\% & 2.88\% & 2.26\% & 2.10\% & 3.52\% & 1.99\% & 2.08\% & 2.96\%\\
    \bottomrule
    \end{tabular}}
  \label{table:performance1}
\end{table*}

\begin{table*}[]
  \centering
  \caption{Experiments comparing our proposed RAM-GNN model with five baseline GNN models in the pre-training step.}
    \resizebox{1\textwidth}{!}{\begin{tabular}{clccccccccccc}
    \toprule
    {Datasets}  
  &\multicolumn{6}{c}{MovieLens}  &\multicolumn{6}{c}{KKBox}  \\
    \cmidrule(lr){1-1} \cmidrule(lr){2-4} \cmidrule(lr){5-7}  \cmidrule(lr){8-10} \cmidrule(lr){11-13} 
   {Models} & HR@10 & MRR@10 & NDCG@10 & HR@20 & MRR@20 & NDCG@20 & HR@10 & MRR@10 & NDCG@10 & HR@20 & MRR@20 & NDCG@20 \\
    \midrule 
      GCN   &  0.148 & 0.049  & 0.070 & 0.218 & 0.051 & 0.079 & 0.727 & 0.516 & 0.543 & 0.803 & 0.498 & 0.522\\
      R-GCN & 0.155 & 0.056  & 0.074 & 0.224 & 0.055 & 0.084 & 0.735 & 0.522 & 0.551 & 0.810 & 0.501 & 0.534   \\
      W-GCN   &  0.153 & 0.057  & 0.076 & 0.220 & 0.057 & 0.086 & 0.730 & 0.523 & 0.546 & 0.807 & 0.505 & 0.531 \\
      VR-GCN  &  0.151 & 0.050  & 0.072 & 0.226 & 0.059 & 0.084 & 0.734 & 0.519 & 0.550 & 0.802 & 0.502 & 0.557 \\
      CompGCN   &  \underline{0.155} & \underline{0.058}  & \underline{0.080} & \underline{0.233}   & \underline{0.064} & \underline{0.101} &   \underline{0.759} & \underline{0.543}  & \underline{0.612} & \underline{0.832}   & \underline{0.553} & \underline{0.624}\\
      RAM-GNN & \textbf{0.164} & \textbf{0.062}  & \textbf{0.087} & \textbf{0.242} & \textbf{0.066} & \textbf{0.107} &  \textbf{0.812} & \textbf{0.584}  & \textbf{0.647} & \textbf{0.873} & \textbf{0.589} & \textbf{0.660}\\
     \midrule
     Improvement & 5.84\% & 6.90\% & 8.97\% & 3.28\% & 3.13\% & 5.94\% & 7.03\% & 7.48\% & 5.72\% & 4.93\% & 6.51\% & 5.71\%\\
    \bottomrule
    \end{tabular}}
  \label{table:performance2}
\end{table*}

\subsubsection{Evaluation Metrics}
Two  evaluation  metrics  are  used to measure the performance of our RAM-GNN framework:

\textbf{HR}. This is the Hit Ratio (HR) of target items that are in the recommendation lists.

\textbf{MRR}. We can measure the performance of the model w.r.t the ranking list of items. Suppose that the model produces a list of items to the user, and the list is ordered by the confidence of the prediction. In this case, a higher MRR score means target items tend to have higher rank positions in the predicted item lists. 

\textbf{NDCG}. This measures the ranking quality. It normalizes the Discounted Cumulative Gain (DCG) to be between 0 and 1 by Ideal Discounted Cumulative Gain (IDCG). 

\subsubsection{Implement Settings}
We implement our model in PyTorch \footnote{https://pytorch.org/}. The number of layers in RAM-GNN is set to 2, and in the recommendation GNN, it is set to 4, according to \cite{wang2019neural}. As will see later, with the help of pre-training, RAM-GNN with a fewer number of layers on a smaller user-item interaction graph outperforms state of the art  GNN based recommendation methods.   

\subsection{Recommendation Performance (RQ1)}
To prove the superiority of our proposed RAM-GNN framework, we conduct experiments on two datasets and compare our model with eight baseline methods. In this section, we denote the pre-training and recommendation steps together as the RAM-GNN framework. 

\subsubsection{Compared Methods}
We compare the performance of our model with the following eight baselines:
\begin{itemize}
    \item \textbf{MF-BPR} \cite{rendle2012bpr}: This is matrix factorization with Bayesian Personal Ranking (BPR) as the loss function.
    
    \item \textbf{FM} \cite{rendle2010factorization}: Factorization Machine (FM) is a content-based model
    which uses feature interactions to model user preferences. We take the side information as additional input features in both datasets.
    
    \item \textbf{NFM} \cite{he2017neural}: Neural Factorization Machine applies an MLP to model the high-order feature interactions. 
    
    \item \textbf{FISM} \cite{kabbur2013fism}: This is an item collaborative filtering (ICF) model which learns the user embedding by aggregating the item embeddings that she has interacted with.

    \item \textbf{CKE} \cite{zhang2016collaborative}: It models the items with a knowledge graph that connects items with their features and then learns the embeddings of items that can be used in CF models.
    
    \item \textbf{LightGCN} \cite{he2020lightgcn}: It simplifies GNN based recommender systems like NGCF \cite{he2017neural} to make them linear models and achieve better performance.
    
    \item \textbf{KGPolicy}\cite{wang2020reinforced}: Knowledge Graph Policy Network is to explore high-quality negatives via reinforcement learning. We take features of users and items as nodes in the knowledge graph.

    \item \textbf{RCF} \cite{xin2019relational}: Relational Collaborative Filtering (RCF)  considers the multi-relational pattern in the items. It utilizes the attention mechanism to combine the multi-relational pattern with the collaborative relation.
    
\end{itemize}

\subsubsection{Results}

The experimental results are presented in Table \ref{table:performance1}. From these results, RAM-GNN clearly outperforms all baselines. We summarize the following observations. 
\begin{itemize}
    \item Compared with RCF, our model achieves better performance on both datasets. RCF is the state-of-the-art recommendation model among all baselines, primarily because it considers the multi-relational pattern in the data. Our model can not only take multi-relational patterns into consideration but also incorporate the graph structure to build the connections at the user-level and item-level.
    \item KGPolicy achieves the best performance compared to all baselines except RCF and RAM-GNN. This proves that the KG can improve the results. However, considering multi-relational patterns is more important than only modeling on the KG because the performance of RCF is better than KGPolicy. Also, KGPolicy may suffer from the over-smoothing problem, which can hinder its performance.
    \item Our model has approximately 10\% improvement over LightGCN, which proves the effectiveness of the pre-training module. Without RAM-GNN in the pre-training step, our model is similar to LightGCN. The pre-training module incorporates the relations between users and items into the recommendation model so that it can be an effective supplement to the collaborative filtering model.
\end{itemize}

\subsection{Pre-training Performance (RQ2)}
To measure the difference of RAM-GNN with other multi-relational GNN models, we compare our model with five state-of-the-art GNN models. In this section, we only substitute the RAM-GNN in the pre-training step with other models but still use the GNN model in the recommendation step. For the models which cannot learn the embeddings of the item features, we only use the item embeddings in the recommendation step.

\subsubsection{Compared Methods}
We compare the performance of our model with the following five baselines:
\begin{itemize}
    \item \textbf{GCN} \cite{kipf2016semi}: This is the widely-used spectral GCN model. Since it is not designed for multi-relational graphs, we only pick one type of relation for each graph in the pre-training.
    
    \item \textbf{R-GCN} \cite{schlichtkrull2018modeling}: Relational GCN enhances the GCN by employing different weight matrices to process multiple types of relations. 
    
    \item \textbf{W-GCN} \cite{shang2019end}: Weighted GCN utilizes an additional trainable relational-specific scalar weight in the GCN model. 
    
    \item \textbf{VR-GCN} \cite{ye2019vectorized}: Vectorized Relational GCN (VR-GCN) learns the embeddings of relations in the GCN framework. Each relation is represented as a vector. 
    
    \item \textbf{CompGCN} \cite{vashishth2019composition}: Composition-based R-GCN (CompGCN) employs KG algorithms in GCN models. The relations are initialized as vectors,  and the KG algorithm (TransE) is applied to learn the embeddings of nodes and relations.
    
\end{itemize}

\subsubsection{Results}
The results comparing pre-training models are shown in  Table \ref{table:performance2}. From the table, we note the following observations:
\begin{itemize}
    \item RAM-GNN outperforms all baseline GNN models in the experiments. This proves the effectiveness of our proposed RAM-GNN. Also, CompGCN outperforms all other baselines on both datasets, indicating that KG algorithms contribute to learning the embeddings.
    \item Our model has a greater improvement on the KKbox dataset than on the MovieLens. This indicates that RAM-GNN has a better capability in handling large graphs. We infer the reason can be the attention layers in RAM-GNN. When the graph becomes larger and denser, the attention mechanism can help to understand the importance of each node and relation, which is proven useful when dealing with complex data \cite{velivckovic2017graph}.
\end{itemize}

\subsection{Ablation Study (RQ3)}
\begin{figure}[tbp]
\centering
\subfigure[MovieLens]{
\includegraphics[width=1.52in]{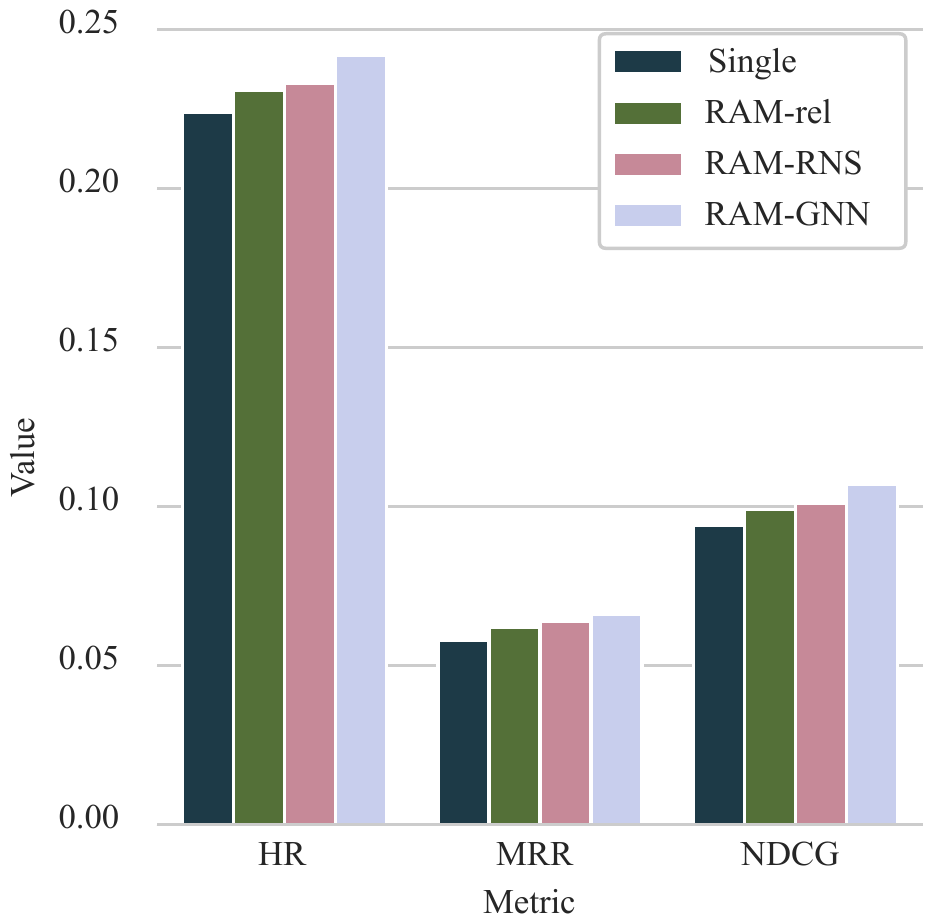}	
}
\quad
\subfigure[KKBox]{
\includegraphics[width=1.52in]{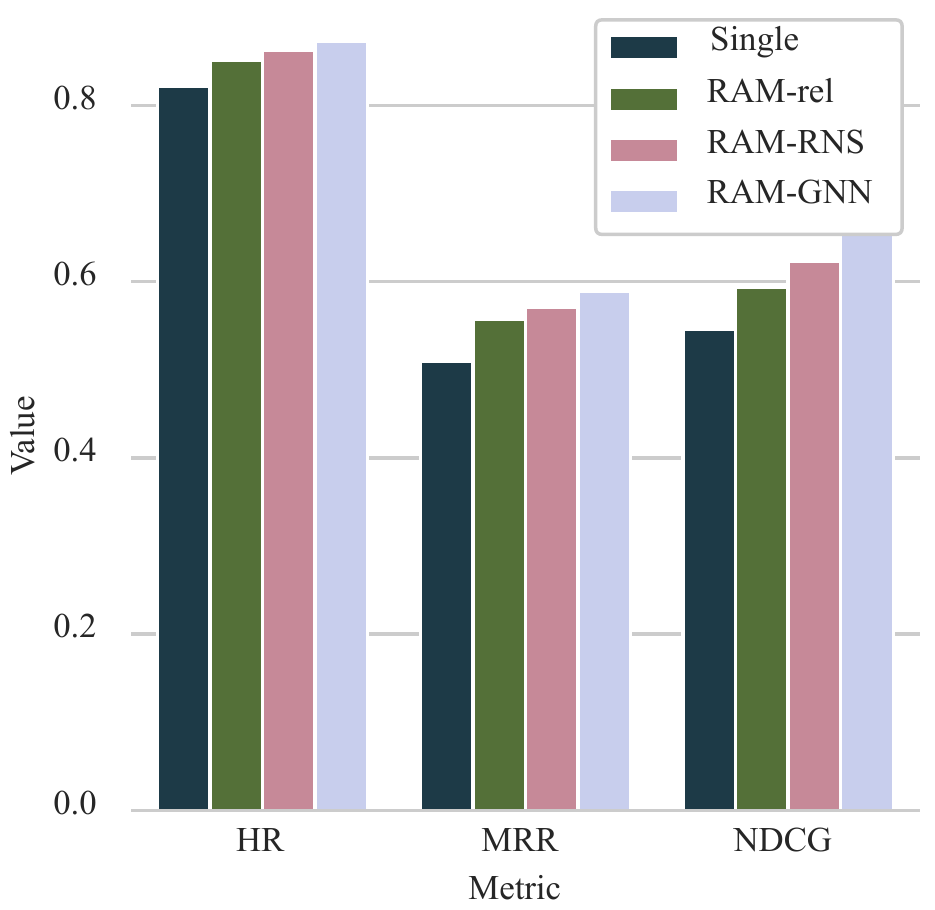}	
}
\caption{Ablation Study on datasets MovieLens and KKBox w.r.t the metrics HR@20, MRR@20, and NDCG@20.}
\label{fig:ablation}
\end{figure}

In this section, we measure the effectiveness of the pre-training, as well as the node and relation-level attention layers. We compare our model with four ablation models: 1) \textit{Single}, which does not the pre-training step. 2) \textit{RAM-RNS}, which only has the reinforced neighbor sampler in the pre-training model. 3) \textit{RAM-rel}, which only has the relation-level attention layer during pre-training, and 4) \textit{RAM-GNN}, which is the complete model we propose. Figure \ref{fig:ablation} displays the comparison between the ablation models. 

From Figure \ref{fig:ablation}, we have the following observations:
\begin{itemize}
    \item Both RAM-RNS and RAM-rel outperform the single model, proving the effectiveness of the node-level and relation-level attention layers. In the best performing RAM-GNN, there can be multiple relations between two entities. If we treat all types of relations equally in the model, it will miss the difference in the meanings of relations. Besides, the node-level attention layer specifies different weights for different nodes in the neighborhood to improve the aggregation quality. Also, the better performance of RAM-GNN compared to a single model shows the importance of pre-training.  
    \item On the KKBox dataset, the performance of RAM-RNS is clearly better than RAM-rel, while their performance is much closer on MovieLens. This phenomenon may result from the complexity of the user and item graph. When the graph is denser, there can be more nodes connected to the central node. With RNS, the model can emphasize  the most similar nodes and reduce redundant information.
\end{itemize}

\subsection{Filtering Threshold Analysis (RQ4)}
\begin{figure}[tbp]
\centering
\subfigure[MovieLens]{
\includegraphics[width=1.53in]{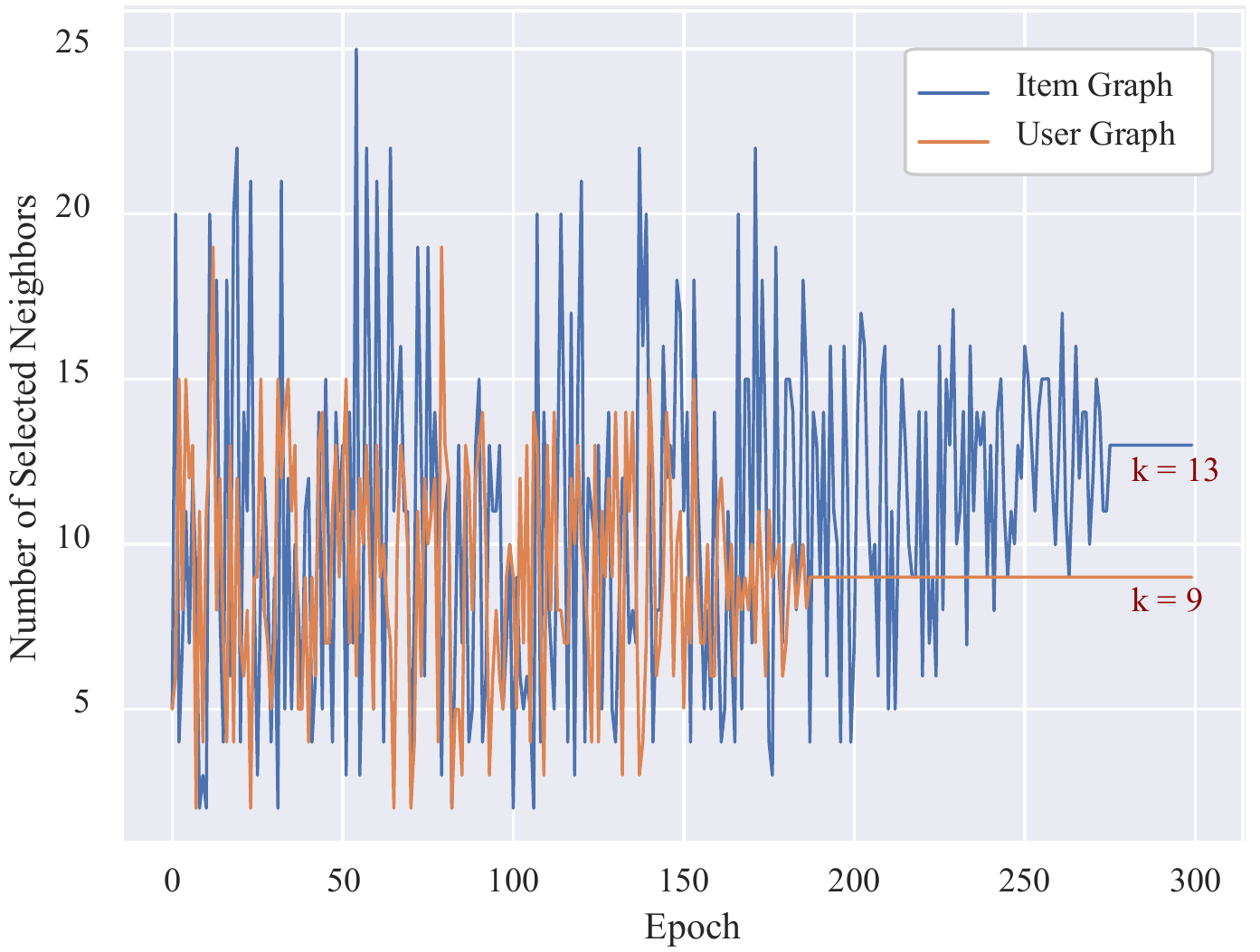}	
}
\quad
\subfigure[KKBox]{
\includegraphics[width=1.53in]{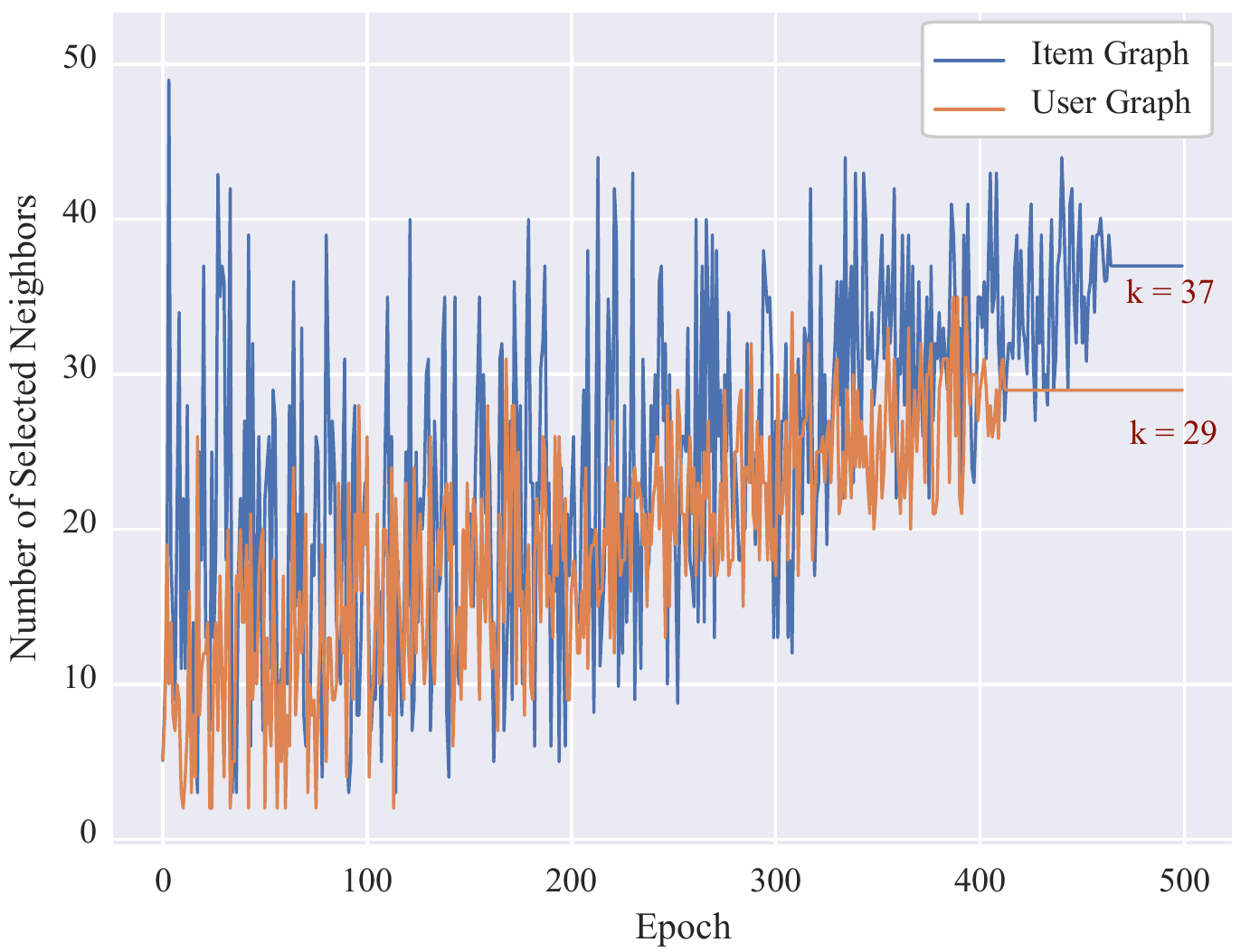}	
}
\caption{The changes of filtering threshold in RL process. We choose user ages and item genres as examples for both datasets. The initial values of both datasets are $5$, and the change of each iteration is $2$}
\label{fig:threshold}
\end{figure}

In this section, we discuss the filtering threshold learned in the RNS module. The experimental results are shown in Figure \ref{fig:threshold}. We choose  user ages and item genres as examples for both datasets. The initial threshold is set as $5$, and $\epsilon$ is $2$. 

From the results in Figure \ref{fig:threshold}, we observe that the filtering thresholds on MovieLens dataset converges when $k=9$ and $k=13$ for the user and item graph, while on KKBox they converges when $k=29$ and $k=37$. Because KKBox has more users and items compared to MovieLens, a larger filtering threshold can ensure the GNN gets enough information in learning the node embeddings.

\subsection{Hyperparameter Analysis (RQ5)}\label{3.5}
In this section, we conduct experiments on two kinds of hyperparameters: composition operations and embedding dimensions. We compare different hyperparameters and summarize the best settings for our model.

\subsubsection{Composition Operations}
We compare the three composition operations mentioned in \hyperref[2.2.1]{Section 2.2.1}. The three operations are addition (add), multiplication (mul), and circular-correlation (corr). Based on the RAM-GNN framework, the experimental results are listed in Table \ref{table:performance3}.

\begin{table}[t]
  \centering
  \caption{Experiments on three composition relations. 'Improv' is the abbreviation of 'Improvements'}
    \resizebox{0.47\textwidth}{!}{\begin{tabular}{clccccccccccc}
    \toprule
    {Datasets}  & {Metrics} & \multicolumn{3}{c}{Operations} & Improv. \\
    \cmidrule(lr){3-5}
    & & add & mul & corr \\
    \midrule 
               & HR@20    & 0.240    & 0.237 & \textbf{0.242}  & 2.07\% \\
    MovieLens \:  & MRR@20   & 0.063 & \textbf{0.068} & 0.066 & 3.03\% \\
               & NDCG@20 \: \: & 0.104  & 0.102 & \textbf{0.107} & 2.88\% \\
     
    \midrule
           & HR@20     & 0.863  & 0.866  &  \textbf{0.873} & 0.80\% \\
    KKBox  & MRR@20    & 0.579  & 0.583  & \textbf{0.589} & 1.03\% \\
           & NDCG@20 \: \:  & 0.657  & \textbf{0.662} &  0.660 &  1.21\% \\ 
    \bottomrule
    \end{tabular}}
  \label{table:performance3}
\end{table}

From the results, we observe that \textit{corr} has the best performance overall. Therefore we utilize \textit{corr} when conducting the experiments in the previous sections. However, the results on these three operations are similar. The largest difference between operations in Table \ref{table:performance3} is only 3.03\%. Therefore, \textit{add} and \textit{mul} are also worth trying during practical use because these operations are simpler to compute, and they can also achieve good performance.

\subsubsection{Embedding Dimensions}
We vary the embedding dimensions from 20 to 100 to see the tendency of the change in predictive performance. The experimental results are displayed in Table \ref{table:dimension}.

From the results, we observe that the optimal embedding dimension is about 60-80. Also, the optimal embedding dimension on KKbox is larger than on MovieLens. This may because the KKbox dataset is much larger than the MovieLens dataset. When training the model, the larger dataset has a stronger ability to optimize embeddings with higher dimensions.

\begin{table}[t]
  \centering
  \caption{Experiments varying the embedding dimension on two datasets.}
    \resizebox{0.47\textwidth}{!}{\begin{tabular}{clccccccccccc}
    \toprule
    {Datasets}  & {Metrics} & \multicolumn{5}{c}{Dimensions} \\
    \cmidrule(lr){3-7}
    & & 20 & 40 & 60 & 80 & 100 \\
    \midrule 
               & HR@20   & 0.2383 & 0.2395 & \textbf{0.2423} & 0.2406 & 0.2414 \\
    MovieLens & MRR@20  &0.0647 &0.0658 &0.0651
&\textbf{0.0663} &0.0655\\
               & NDCG@20 & 0.1052 &0.1043 &\textbf{0.1071} &0.1065 &0.1058\\
     
    \midrule
           & HR@20   &0.859 &0.861 &0.865 &\textbf{0.873} &0.863\\
    KKBox  & MRR@20  &0.571 &0.575 &0.583 &0.581
& \textbf{0.589} \\
           & NDCG@20 & 0.651 &0.655 &0.657 &\textbf{0.661} &0.659\\ 
    \bottomrule
    \end{tabular}}
  \label{table:dimension}
\end{table}

%% file: Sections/related-works.tex
\section{Related works}
\subsection{GNN-based Recommendation Models}
GNNs have proven to be useful in different areas \cite{peng2019fine, liu2021medical, dou2020enhancing, liu2020heterogeneous, cao2021knowledge, li2021higher, hei2021hawk}. There also exists a rich literature utilizing the graph structures in data to provide recommendations. Among them, there are two main directions about which graph to use. One direction is to use the user-item bipartite graph to derive recommendations. Among them, \cite{zheng2018spectral, berg2017graph, yu2020graph} directly perform convolution operations to explore interactions between users and items. \cite{wang2019neural, he2020lightgcn} leverage layer-to-layer aggregation functions to capture the high-order connections. \cite{li2021dynamic} models user-item interactions with a dynamic graph. These methods apply GNN on the user-item interaction graph from different aspects. The GNN structure has advantage of representing high-dimensional graph data into low-dimensional embeddings without feature engineering, so it is suitable to be directly implemented on the user-item interaction graph. However, the interaction graph ignores the rich features of users and items.

The other direction is to use the KG to provide recommendations.  This direction contains two different categories of approaches as well. One category is non-GNN models. Many early studies~\cite{luo2014hete,hu2018leveraging,xian2019reinforcement,wang2018dkn,xin2019relational} derive embeddings from KGs via optimization methods and utilize the embeddings in downstream collaborative filtering/prediction steps. These methods are more efficient than recent GNN-based approaches, since they only use the KG embeddings as auxiliary information. However, they tend to be less effective because they fail to incorporate KG information into the end-to-end architecture. The other category is GNN-based models, which tend to apply GNN on KG to derive representations for all nodes (and even relations)~\cite{wang2018ripplenet,wang2019knowledge,tang2019akupm,zhao2019intentgc}. Besides, \cite{liu2020basket, chang2020bundle, liu2020basconv} construct a heterogeneous graph containing users, items and baskets as three kinds of nodes. These approaches integrate information from both the collaborative bipartite graph and the user and
item features, and thus achieve better performance. However, dealing with different relations is still challenging for these models, since GNNs are naturally designed for homogeneous graphs.


\subsection{Multi-Relational Graph Neural Networks}
Currently, Graph Neural Networks (GNNs) have been widely explored to process graph-structured data. Motivated by convolutional neural networks, Bruna et al. \cite{estrach2014spectral} propose graph convolutions in the spectral domain. Then, Kipf and Welling~\cite{kipf2016semi} simplified the previous graph convolution operation and designed a Graph Convolutional Network (GCN) model. To inductively generate node embeddings, Hamilton et al. proposed the GraphSAGE~\cite{hamilton2017inductive} model to learn node embeddings with sampling and aggregation functions. All these models have demonstrated their superior performance on many tasks, e.g., link prediction and node classification.

When the graph has multiple kinds of relations~\cite{wang2021dskreg} between a pair of nodes, it forms a more complex graph structure, and we call it multi-relational graph. There are some traditional methods handling multi-relational graph including TransE, TransR, DistMult. TransE \cite{bordes2013translating} embeds entities and relations following the translational principle. TransR \cite{lin2015learning} extends TransE by separating different spaces for entities and spaces. DistMult \cite{yang2014embedding} represents relations as diagonal matrices. These methods have competitive performance on different tasks, but have limitation in recovering missing facts in the multi-relational graph.

After the appearance of GNNs, many studies extended them to deal with multi-relational graph. Relational GCN \cite{schlichtkrull2018modeling} is the first work that learns node embeddings from multi-relational graphs. For each kind of relation, it utilizes a weight matrix to represent the relation. It demonstrates that R-GCN is able to handing the missing fact issue. Based on this work, weighted GCN \cite{shang2019end} adds a trainable weight on each relation in the GCN model. Vectorized Relational GCN (VR-GCN) \cite{ye2019vectorized} learns node embeddings as well as relations embeddings with the knowledge base algorithm TransE \cite{bordes2013translating}. Composition-based Relational GCN (CompGCN) \cite{vashishth2019composition} is more flexible because it can apply knowledge base algorithms when learning node and relation embeddings. All the models above have a common limitation in that they cannot deal with the graphs that have several relations between two nodes. 


%% file: Sections/conclusion.tex
\section{Conclusion}
In this paper, we introduce a model, Reinforced Attentive Multi-relational Graph Convolutional Network (RAM-GNN), to  pre-train the user graph and item graph. RAM-GNN has two main modules. First is relation-level attention which calculates the  importance of different types of relations. The second one is Reinforced Neighbor Sampler (RNS), which searches the most similar neighbors iteratively to reduce redundant information and improve efficiency. From the pre-training step, we learn the user and item embeddings. With another GNN for the user-item interaction graph, we inject the knowledge learned from pre-training to the collaborative filtering model and then make recommendations. Experimental results show that our proposed RAM-GNN also has the best performance, as compared to other multi-relational GNNs.

On top of the relations we present in this work, there are some other potential relations among users and items to be explored in the pre-training step, e.g. time dependency of interactions and social relations in users. In the future, we will try to extend our model on sequential recommendation and social recommendation to utilize more kinds of relations and improve performance.

\section{Acknowledgement}
The authors of this paper were supported by the S\&T Program of Hebei through grant 21340301D,  in part by NSF under grants III-1763325, III-1909323, III-2106758, and SaTC-1930941.